\documentclass[
preprint
]{elsarticle}

\usepackage{graphicx}
\usepackage{bm}
\usepackage{amsmath}
\usepackage{amssymb}
\usepackage{bigstrut}
\usepackage{multirow}

\newcommand{\Ref}[1]{Ref.~\cite{#1}}   
\newcommand{\bea}{\begin{eqnarray}}
\newcommand{\eea}{\end{eqnarray}}
\newcommand{\beq}{\begin{equation}}
\newcommand{\eeq}{\end{equation}}

\newcommand{\Eq}[1]{Eq.\,(\ref{#1})}
\newcommand{\ket}[1]{\vert{#1}\rangle}

\begin{document}



\title{Comparison of different eigensolvers 
for calculating vibrational spectra
using low-rank, sum-of-product basis functions. }

\author[metz]{Arnaud Leclerc}
\ead{Arnaud.Leclerc@univ-lorraine.fr}

\author[kingston]{Phillip S. Thomas}
\ead{Phillip.Thomas@chem.queensu.ca} 

\author[kingston]{Tucker Carrington \corref{corres}}
\ead{Tucker.Carrington@queensu.ca}

\address[metz]{Universit\'e de Lorraine, UMR CNRS 7565 SRSMC, \\ 1 boulevard Arago 57070 Metz, France}
\address[kingston]{Chemistry Department, Queen's University, Kingston, \\ Ontario K7L 3N6, Canada}

\cortext[corres]{Corresponding author}

\date{}





%

\begin{abstract}

Vibrational spectra and wavefunctions of polyatomic molecules 
can be calculated at  low memory cost
using low-rank sum-of-product (SOP)  decompositions to represent  basis functions generated using an iterative eigensolver.
Using a  SOP tensor  format does not determine  the iterative eigensolver. 
The  choice of the interative eigensolver  is limited by the need to restrict the rank of the SOP   basis functions at every stage of the calculation. 
We have adapted, implemented and compared different reduced-rank algorithms based on 
standard iterative methods (block-Davidson algorithm, Chebyshev iteration) 
to calculate vibrational energy levels and wavefunctions of the 12-dimensional acetonitrile molecule.  
The effect of using low-rank SOP  basis functions on  the different methods is analyzed 
and the numerical results are compared with those  obtained with  the reduced rank block power method introduced in J. Chem. Phys. 140, 174111 (2014). 
Relative merits of the different algorithms are presented, showing that the advantage 
of using a more sophisticated method, although mitigated by the use of reduced-rank 
sum-of-product functions, is  noticeable in terms of CPU time.

\end{abstract}

\maketitle

\section{Introduction}



Understanding the internal motion of the nuclei of a polyatomic molecule is an important problem in molecular physics. 
Given a potential energy surface, the most general approach to compute vibrational spectra is to calculate eigenvalues and  eigenvectors of a matrix 
  representing the Hamiltonian operator in some basis.  When a direct product basis, with functions that are products of functions of a single coordinate, is used,
the computational cost 
 increases exponentially with the number of internal degrees of freedom ($D=3N-6$ for a molecule with $N$ atoms). This problem  is well-known as the ``curse of dimensionality". 
For example, to use a direct product basis to calculate the   vibrational spectrum for a molecule with 6 atoms,
one would need  about   
$8 \times 10^{15}$ GB
to store the matrix, assuming there are 10 basis functions for each coordinate.   
If one uses an iterative eigensolver   \cite{lanczos,cullum,davidson1975,liu1978,ribeiro2005}
it is possible to compute a spectrum without storing (or computing) the Hamiltonian matrix
\cite{bramley1993_2,bowman2008,bramley1994,neuhauser1994,iung1995,wang2003_2}. 
One must only store a few vectors, however a single vector requires about   8000 GB.   
Iterative methods only require the ability to apply the Hamiltonian matrix  to vectors.

Different strategies have been  employed to drastically  reduce the memory cost of storing these vectors.  To some extent they can be used together.   
 These include: 
pruning the basis set by retaining only some of the direct product basis functions 
\cite{dawes2005,avila2011,lauvergnat2013,halverson2015,robert2016,avila2009,halonen1983,bowman2003,wang2001};
using basis functions localized in the classically allowed regions 
\cite{davis1979,halverson2012,james6d,shimshovitz2012}; 
making contracted basis functions by diagonalizing reduced-dimension Hamiltonians for 
 strongly coupled coordinates \cite{carter1988,bramley1993,wang2002}.
All of these strategies aim to reduce the number of required basis functions.  
 Instead, one can work with a direct product basis, but obviate the need to 
store vectors with $n^D$ components,  where $n$ is a representative number of basis functions for a single coordinate.   This can be achieved by representing basis vectors 
with  low-rank  tensors.   One can think of the multi-configuration time-dependent Hartree  (MCTDH) method as representing wavefunctions in the Tucker tensor format
\cite{mctdhbook,mctdhrev,kolda2009}. 
The Tucker format does not  defeat the curse of dimensionality because it is a direct product representation.   The advantage of MCTDH is the optimization of the 1D basis functions.
There is also a multi-layer MCTDH method using what mathematicians call a  hierarchical-Tucker format \cite{wang2003,manthe2008}. 
There are several other tensor formats that can be used,   such as the Matrix Product States \cite{schollwock2011} (equivalent to the Tensor Train format \cite{oseledets2011,rakhuba2016}) 
or the Canonical decomposition (CP format) \cite{kolda2009,beylkin2005},  which  we exploit in this article.

Tensors can be used to compute vibrational energy levels in two ways.   
One way to use tensors 
 is to represent the desired eigenvectors in tensor format and to optimize elements of the tensors. 
  The well-known density matrix renormalization group (DMRG)  method is of this type.  
Another way is to compute eigenvalues by projecting into a space each of whose vectors is in a  tensor format.
The vectors may be calculated using a 
 standard iterative algorithm. In this article we use this second way. 
It can only be used  if   the   Hamiltonian matrix is also low rank.
The tensor format we use is CP and we therefore require that the   Hamiltonian be    a sum of products.   
Whenever the Hamiltonian is applied to a vector, the rank (i.e. the number of terms)  increases and must   be reduced to keep the memory cost within acceptable limits.
This reduction step also requires optimization and is the computational bottleneck.
For both ways,   the optimization is usually  done  with a variant of the     alternating least squares algorithm. 
The idea of using a basis of vectors in a tensor format to compute vibrational levels was introduced in \Ref{leclerc2014}, where the 
 Reduced Rank Block Power Method (RRBPM) was presented.   It was later shown that  the RRBPM  is    compatible with symmetry partitioning of the basis set,
which   facilitates computing and  assigning  levels \cite{leclerc2016}.   
A  hierarchical version of the RRBPM,  taking advantage of successive contractions for strongly coupled coordinates was proposed in \Ref{thomas2015}. 
Making intermediate basis functions using a tree structure greatly improved the accuracy and reduced the required CPU time. 
The H-RRBPM of \Ref{thomas2015} uses the two-layer RRBPM of \Ref{leclerc2014} at each layer.

In this article, we test a method similar to the original RRBPM but with a better eigensolver. We do not use the hierarchical version, but the ideas we introduce could be used in conjunction with it.   The power method used in \cite{leclerc2014} is a crude  and simple iterative method. It is well known to  converge slowly, especially if the density of states is high.  
The CPU time required to obtain converged eigenvalues can thus be quite long.
The power method was originally used not only because it is simple but also because the intermediate vectors it generates   become increasing similar to eigenvectors which are assumed
to be ``reducible".    Here reducible means that little error is incurred when the rank is reduced.
In this article, we show that it is possible to use more elaborate  eigensolvers. 
In section \ref{theory}, 
we briefly recall the main ideas of the RRBPM   and introduce the modifications required to use  two other eigensolvers.     One is a  Chebyshev  filter 
method and the other is  a  block-Davidson method. 
These algorithms are then used to calculate vibrational eigenstates of  acetonitrile  (CH$_3$CN), a 12D problem, in section \ref{results}.


\section{Reduced-rank Iterative Methods for calculating spectra \label{theory}}


\subsection{Sum-of-product format for functions and operators \label{secsopformat}}

In all calculations  in this article basis vectors  are in   CP format \cite{kolda2009}.  Eigenvectors of the Hamiltonian matrix are obtained as linear combinations of the 
CP basis vectors and hence they also are in CP format.    
In general, one can expand  a multidimensional function  in a  direct product basis,
\begin{equation} 
F(q_1, \dots, q_D) \simeq \sum_{i_1=0}^{n_1-1} \dots \sum_{i_D=0}^{n_D-1} F_{i_1 i_2\dots i_D}  
\prod_{k=1}^D \theta_{i_k}^k (q_k),
\label{wavefunction}
\end{equation}
where $\{ \theta_{i_k}^k (q_k) $, $i_k=0\dots(n_k-1) \}$, is a set of basis functions associated with coordinate $q_k$, $k=1\dots D$. 
The expansion coefficient is in CP format if 
\begin{equation}
F_{i_1 i_2 \dots i_D}  = \sum_{\ell=1}^R  \prod_{k=1}^D f^{(\ell,k)}_{i_k} ~,
\label{sop}
\end{equation}
where the $f^{(\ell,k)}_{i_k}$, $i_k=0\dots(n_k-1)$ are components of one-dimensional vectors ${\bf f}^{(\ell,k)}$ which 
generally
appear only once each in the expansion.  There is no need for them to 
be orthogonal or normalized.
If   $  F_{i_1 i_2\dots i_D}   $  is in CP format then  $F(q_1, \dots, q_D)$ is a sum of products (SOP),
\begin{equation}
F(q_1, \dots, q_D)=\sum_{\ell =1}^R \prod_{k=1}^D \phi^{(\ell,k)} (q_k)  
=\sum_{\ell =1}^R \prod_{k=1}^D \left( \sum_{i_k=0}^{n_k-1} f^{(\ell,k)}_{i_k} \theta_{i_k}^k(q_k) \right) ~. 
\label{Fexpansion}
\end{equation}
The important idea underlying all tensor-based methods is:  in general  the memory cost of storing 
 $F_{i_1 i_2 \dots i_D}$ scales as $n^D$ but the memory cost of storing the right side of \Eq{sop} scales as  $RDn$
  \cite{leclerc2014,thomas2015}.


Vectors generated by the iterative eigensolver from a vector in CP format will themselves only be in CP format if the Hamiltonian is a SOP.
Throughout this article, we shall  assume that the potential energy surface is known and expressed  as a SOP.
The kinetic energy operator (KEO)  is often a SOP.  In this article we use normal coordinates and neglect the $\pi-\pi$ term \cite{watson1968}
 so that the KEO is a sum of terms each of which depends on 
a single coordinate,                   
\begin{equation}
\hat{H}(q_1,\dots,q_D) = \sum_{k=1}^T \prod_{j=1}^D  \hat{h}_{k j} (q_j),
\label{Hamiltoniansop}
\end{equation}
where $\hat{h}_{kj}$ is a one-dimensional operator acting in a Hilbert space associated with coordinate $q_j$.
The application of $\hat{H}$ on vectors is the crucial step in every iterative method. 
When vectors are in CP format and the operator is a SOP,   the application of {$\bf{H}$}
 requires only one-dimensional operations, this can be seen by multiplying Eq. \eqref{Hamiltoniansop} and Eq. \eqref{Fexpansion} \cite{leclerc2014}. Only the small $(n_j \times n_j)$ Hamiltonian matrices representing $\hat{h}_{k j}(q_j)$ have to be computed and stored in memory. 
In the following  subsections, we describe several methods to make a small basis of  low-rank SOP basis functions with which one can compute low-lying levels of molecules with more than five atoms.

\subsection{Reduced-rank Block Power method \label{RRBPM}}

In the power method, the matrix ${\bf H}$   is applied recursively to a start  vector ${\bf F}_0$,  to calculate  the eigenvector associated with the largest
 eigenvalue \cite{saad2011,strang1986}. To obtain several eigenvalues one uses  a block of vectors.    To calculate the lowest eigenvalues, the 
 matrix is also shifted  by $-\sigma  {\bf I}$.    The RRBPM uses  a shifted block power method and stores   ${\bf H}$  as a SOP of small matrices and the vectors in CP format.  
Each matrix-vector product increases the rank of a vector by a factor of $T$.   
 The rank also increases, to a lesser extent,   when vectors in the block are orthogonalized and updated, see \Ref{leclerc2014}.
The rank   must be reduced after each operation which increases it. 
As in Ref. \cite{leclerc2014,leclerc2016}, an old (large-rank) vector
  ${\bf F}^{\text{old}}$ is reduced from rank $R^{\text{old}}$ (typically a few thousands) to rank $R^{\text{new}}$ (a few tens) using an alternating-least-squares (ALS) algorithm. 
For consistency of comparison we will use the same implementation of  ALS as in \cite{leclerc2014,leclerc2016}, based on \cite{beylkin2005}.  ALS is an iterative process in which linear equations are solved to find new vectors  
 $^{\text{new}}f^{(\ell,k)}_{i_k}$ which minimize the difference $ \parallel {\bf F}^{\text{new}} - {\bf F}^{\text{old}} \parallel $. 
The ALS reduction is a crucial step in the reduced rank methods described below and is responsible for a majority of the calculation time. The main reduction has to be performed after the matrix-vector product which is responsible for the most important increase of the rank.
 The algorithm can be summarized as follows:

\begin{enumerate}
\item Generate initial guess for the  block of SOP eigenvectors $\bm{\mathcal{F}}=({\bf F}_{k=1}\dots{\bf F}_{k=B})$ 
\item For $m=1 \dots N_{rs}$ (with $N_{rs} \simeq 10$)   
	\begin{enumerate}
		 \item For $k=1\dots B$, ${\bf F}_k^{m+1} \leftarrow ({\bf H}-\sigma {\bf I}  ) {\bf F}_k^m $. 
	     \item Reduce the ranks using alternating least squares.  
	\end{enumerate}
\item Orthogonalize the vectors, make a matrix ${\bf H}'= \bm{\mathcal{F}}^t \bf{H} \bm{\mathcal{F}}$ representing $\hat{H}$  in this  SOP  basis set and compute the overlap matrix ${\bf S}=\bm{\mathcal{F}}^t \bm{\mathcal{F}}$.
\item Solve the generalized eigenvalue problem  to obtain eigenvalues and eigenvectors.    
\item Reduce the rank, update the vectors with eigenvectors; go back to step 2. 
\end{enumerate}
In the above algorithm $N_{rs}$ is the number of power iterations done between two updates of the basis vectors using reduced eigenvectors of ${\bf H}'$. 
Note that  the basis size does not grow during the calculation; in step 2 (a) 
   the basis  at   step $(m+1)$ replaces the basis at step $m$.
The energy shift is $\sigma = (E_B+E_{max})/2$, where $ E_{max}$ is the highest eigenvalue of $\bf{H}$
 determined by  doing  a few unshifted power iterations \cite{leclerc2014}.
The memory cost scales as 
\beq
\mathcal{O}(BTRDn)
\eeq
where $B$ is the size of the computed subspace, $T$ the number of terms in the Hamiltonian, $R$ the reduction rank (we keep it  fixed for simplicity), $D$ the number of coordinates and $n$ the number of basis functions per coordinate. 
The cost of performing matrix vector products in the subspace scales as 
\beq 
\mathcal{O}(N_{\text{pow}}BTRDn^2)
\label{eq:costmvp}
\eeq 
where $N_{\text{pow}}$ is the number of power iterations. The overall computational cost of rank reductions using ALS \cite{beylkin2005} scales as 
\beq
\mathcal{O}(N_{\text{pow}} N_{\text{ALS}} B D (R^3 + nTR^2))
\label{eq:costals}
\eeq
with $N_{\text{ALS}}$ as the number of internal iterations for the least squares procedure used to  reduce  the rank. The memory cost and the computational cost of reductions thus scale linearly with $D$.

\subsection{Reduced-rank Chebyshev filter  technique \label{chebyshev}}

There are many ways to push a block of vectors towards the space spanned by the eigenvectors with the lowest eigenvalues. 
The shifted block power method does push a block of vectors in the right direction, however, the space it generates includes 
contributions from eigenvectors whose corresponding eigenvalues are not in the set of the lowest $B$. 
It might be better to use a filter that excludes vectors outside the block.   
Here, we apply a sequence of filter functions to each vector in the block, where a different filter is used for each vector.
The filter for the $k^{th}$ vector is designed to exclude all vectors corresponding to eigenvalues above the $k^{th}$. 
The filters we use are carefully chosen polynomials,  
$\mathbf{F}_{k,purified}=P^k_{m}\left(\mathbf{H}\right)\mathbf{F}_k$,
where $P^k_{m}$ is a  polynomial of degree $m$.

Following Saad \cite{saad1984,saad2011}, we  use a  Chebyshev polynomials of the first kind, 
$P^k_{m}\left(\mathbf{H}\right)=T_{m}\left(\tilde{\mathbf{H}}_k\right)$,
where 
$\tilde{\mathbf{H}}_k=\frac{  \mathbf{H}-d_k \mathbf{I}     }{c_k}$ 
denotes a Hamiltonian that has been shifted and scaled.  We choose  the scaling parameter $c_k$ and the shift parameter $d_k$ as
\begin{align}
c_{k}&=\frac{E_{max}-E_{k+1}}{2} \nonumber\\
d_{k}&=\frac{E_{max}+E_{k+1}}{2} \label{eq:scaling} ~. 
\end{align}
This choice means that for $ \mathbf{F}_k$,
  the contribution of eigenvectors corresponding to   the   eigenvalues  $E_{k+1}\ldots E_{max}$, 
which are mapped into  the interval $\left[-1,1\right]$, 
is decreased
and   the contribution of eigenvectors corresponding to    the  eigenvalues $E_{1}\ldots E_{k}$,
 which   are mapped to $<-1$,
 is increased  \cite{saad1984,saad2011}.
This is true because  
if    $-1\leq x\leq1$,   $-1\leq T_{m}\left(x\right)\leq1$
and  if  $x$ is outside of $\left[-1,1\right]$, $T_{m}\left(x\right)$ is the most rapidly growing of all $m^{th}$-order polynomials. 
This filter will therefore accentuate  contributions from eigenvectors whose eigenvalues are less than $E_{k+1}$.
Using a different filter for each vector in the block makes the vectors more reducible.
Each filter accentuates contributions from eigenvectors with smaller eigenvalues more than eigenvectors with larger eigenvalues.
For instance, if there are $B=80$ states in the block, the filter for the $B^{th}$ vector accentuates the $E_{0}$ eigenvector more than
the ${E_{B}}$ eigenvector. As a result, it is necessary to orthogonalize the $k^{th}$ vector to vectors $1 \ldots k-1$ after applying the filter.

To compute parameters $c_k$ and $d_k$, one needs an estimate of eigenvalues $E_{k+1}$ and $E_{max}$. 
We estimate $E_{max}$ by doing a few (non-shifted) power iterations. 
If the $E_{max}$  estimate is too low, then the true value of $E_{max}$ will be mapped to a value larger than one 
 and vectors in the block will  become ``contaminated''  with contributions from the eigenvector corresponding to $E_{max}$, ruining convergence. 
To avoid this we ``pad'' our estimate of $E_{max}$ by adding $0.01(E_{max}-E_{min})$.
For $k = B$, we do not have an estimate of $E_{k+1} = E_{B+1}$ since $E_{B+1}$ lies outside of the block. In this case we use the estimate $E_{B+1}\approx 2E_{B}-E_{B-1}$  where $E_B$ and $E_{B-1}$ are from the previous iteration.
If the actual spacing between $E_{B}$ and $E_{B+1}$ is smaller than the spacing between $E_{B}$ and $E_{B-1}$, $E_{B}$ will converge slowly.

To apply the  polynomial filter $P_{m}\left(H\right)$  one must evaluate  $m$ matrix-vector products.  This is done recursively. 
  Beginning with $\mathbf{F}_k^{(0)}\equiv\mathbf{F}$ and $\mathbf{F}_k^{(1)}=\tilde{\mathbf{H}}_k\mathbf{F}_k^{(0)}$, the other filtered vectors are obtained from   
\begin{equation}
\mathbf{F}_k^{(j)}=2\tilde{\mathbf{H}}_k\mathbf{F}_k^{(j-1)}-\mathbf{F}_k^{(j-2)}\qquad\left(j=2\ldots m\right)
\end{equation}
Augmenting the degree of the filter by one requires   one additional 
 matrix-vector product and  one vector-vector addition.   After    $  \mathbf{F}_k^{(j)} $   is generated, it is reduced with    ALS. 
Each matrix-vector product is done by exploiting the CP format of the vector.  
As explained by Saad \cite{saad1984,saad2011}, the polynomial filter can magnify the coefficient of the ground state eigenvector beyond the overflow limit if high Chebyshev orders are used. The use of a scaling parameter $\sigma_{j}$ is necessary to prevent this.
The algorithm is:

\begin{enumerate}
\item Generate an initial  block of   vectors $\bm{\mathcal{F}}=({\bf F}_{1}\dots{\bf F}_{B})$, initial 
eigenvalue estimates, and an initial spectral range. 
\item For $k=1 \dots B$: 

\begin{enumerate}
\item Set $c_k$ and $d_k$ as described in Eq. (\ref{eq:scaling}); set $\sigma_{1}=\frac{c_1}{d_1-E_{1}}$; compute $\mathbf{F}_k^{(1)}=\sigma_{1}\tilde{\mathbf{H}}_k\mathbf{F}_k^{(0)}$
\item For $j=2\ldots m$: 

\begin{enumerate}
\item Compute $\sigma_{j}=\left(\frac{2}{\sigma_{1}}-\sigma_{j-1}\right)^{-1}$
\item Chebyshev iteration: $\mathbf{F}_k^{(j)}=\sigma_{j}\left(2\tilde{\mathbf{H}}_k\mathbf{F}_k^{(j-1)}-\sigma_{j-1}\mathbf{F}_k^{(j-2)}\right)$ 
\item Reduce the rank of $\mathbf{F}_k^{(j)}$ using ALS
\end{enumerate}
\end{enumerate}
\item Orthogonalize the vectors, make the matrix ${\bf H}'=\bm{\mathcal{F}}^{t}{\bf {H}\bm{\mathcal{F}}}$
and overlap matrix ${\bf S}=\bm{\mathcal{F}}^{t}\bm{\mathcal{F}}$. 
\item Solve the generalized eigenvalue problem  to obtain eigenvalues and eigenvectors; 
update vectors in $\bm{\mathcal{F}}$; reduce ranks; 
back to step 2.  
\end{enumerate}

The computational cost of filtering a subspace of dimension $B$ is dominated by the cost of evaluating $m$ matrix-vector products and doing $m$ rank reductions, i.e. 
\beq
\mathcal{O}(mBTRDn^{2}) + \mathcal{O}(mN_{\text{ALS}}BD(R^{3}+nTR^{2})).
\eeq
We typically use polynomials of degree $m=10$.

\subsection{Reduced-rank Block Davidson method \label{RRBDM}}

The original Davidson algorithm begins with a start vector and  builds  a space adapted to the calculation of a single eigenvalue  by adding one vector at a time \cite{davidson1975}. 
 The best 
estimate of the desired eigenvalue is obtained by projecting the eigenvalue problem  into the space spanned by the Davidson vectors.  
     Davidson uses a form of preconditioning to favor the convergence of the desired eigenvalue.  We use a block version of Davidson \cite{liu1978}. 
 From one iteration to the next the power method replaces the previous basis with a new basis of the same size.   The size of the matrix does not increase as the calculation proceeds.
 The size of the block Davidson basis does  increase  during the iteration because at each iteration $B$ vectors are added to the basis.  
$B$ new vectors are generated and then orthogonalized with respect to vectors already in the basis.   
The basis is then augmented with the orthogonalized vectors, a generalized eigenvalue problem is solved 
and eigenvectors with the lowest eigenvalues are used either to compute the $B$ new vectors to be added at the next iteration (see algorithm), or  as new start vectors.  
The block Davidson algorithm  we use is essentially a CP version of the one in \Ref{liu1978,ribeiro2005}.   
We restart the algorithm every $N_{rs}$ iterations.
The algorithm to compute $B$ eigenvectors is:

\begin{enumerate}
\item Preparation: 
\begin{enumerate}
\item Define an initial subspace $\bm{\mathcal{F}} = ( {\bf F}_1  \dots  {\bf F}_B )$; set $B'=B$.
\item Make the matrix ${\bf H}'= \bm{\mathcal{F}}^t \bf{H} \bm{\mathcal{F}}$ and overlap matrix ${\bf S}=\bm{\mathcal{F}}^t \bm{\mathcal{F}}$. 
\item Solve the eigenvalue problem, reduce the ranks of the eigenvectors. 
\item Select $B$ eigenvectors ${\bf{\psi_m}}$ corresponding to the lowest eigenvalues $E_m$, $m=1\dots B$. 
\end{enumerate}

\item For $j=1 \dots N_{cycle} $:

 \begin{enumerate}

 \item For $k=1 \dots N_{rs}$: 

   \begin{enumerate}

   \item Compute the residuals ${\bf q}_m = ({\bf H}-E_m {\bf I})  {\bf \psi}_m $; reduce the rank.  \label{for2}
   \item Precondition: Compute (in an approximate way, see text below) the new vectors \\ ${\bf F}_{B'+m} = (E_m - {\bf H}_0 )^{-1} {\bf q}_m $ for $m=1 \dots B$. \label{alg:precond}
   \item Orthogonalize with respect to previous vectors in $\bm{\mathcal{F}}$, reduce the rank. \label{step:orth}
   \item Set $B' \leftarrow B'+ B$; augment ${\bf H}'$;  
augment ${\bf S}$;

   \item Diagonalize, reduce the ranks of the eigenvectors. \label{diag}
   \item Select $B$ eigenvectors ${\psi}_m$ \label{select} corresponding to the lowest eigenvalues $E_m$, $m=1\dots B$. 

   \end{enumerate}
   
   \item Restart: set $B'=B$, update the basis set, keeping only the first $B$ eigenvectors approximations; update ${\bf H}'$ and ${\bf S}$.  
 
 \end{enumerate}
 
\end{enumerate} 
A good starting block is important for the convergence of the Davidson algorithm. We use rank-one eigenvectors of the uncoupled, separable part of the Hamiltonian as start vectors. 

The main difficulty is applying the preconditioner in step~(\ref{alg:precond}). 
In our calculations   ${\bf H}_0$ is diagonal.   The corresponding operator is the separable part of the Hamiltonian.   
With this choice, ${\bf H}_0$ is naturally represented as a sum of products.
$ {   ({\bf  E}_m - {\bf H}_0 )^{-1} }$ is also diagonal
in the direct product basis. 
However, it is not  in  low-rank sum-of-product form.     
The corresponding operator  
 can be written as a sum of $\prod_k n_k$ terms of rank one
by using a spectral expansion of $(E_m - {\bf H}_0 )^{-1}$ in the direct product basis set, 
\beq
\sum_{i_1\dots i_D} \frac{1}{E_m-E^0_{i_1\dots i_D}} \ket{ \Theta_{i_1 \dots i_D} } \langle  \Theta_{i_1 \dots i_D} \vert
\label{precond1}
\eeq
where $\{E^0_{i_1\dots i_D}, \Theta_{i_1 \dots i_D} \}$ are the eigenpairs of ${\bf H}_0$.
We do not use  ${   ({\bf E}_m - {\bf H}_0 )^{-1} }$ because applying it to a vector would increase it rank by a factor of  $\prod_k n_k$.  
Instead, we replace diagonal elements of $    ({\bf E}_m - {\bf H}_0 )^{-1} $  with  
$E_{0,\text{cut-off}}$,  if the 
${i_1 \dots i_D}$ diagonal element of    ${\bf {H}_0  }$  is larger than  $E_{0,\text{cut-off}}$.  
Denote  this modified matrix  $ {\bf M}_{J_0}$, where  $J_0$  is 
the set of indices, 
${i_1 \dots i_D}$, for which  $E_{0,   {i_1 \dots i_D}}  \le     $   $E_{0,\text{cut-off}}$.    There are $N_{lr}$ ($lr$ means ``low rank")  elements in     $J_0$.  
    ${\bf M}_{J_0}$ is a matrix whose rank is 
 $\prod_k n_k$.   
 It can be written as a sum of two matrices,
\beq
{\bf M}_{J_0} =  {\bf M}_{J_0}^{lr} + \lambda {\bf I}
\eeq
 where 
$\lambda = \frac{1}{E_m-E_{0,\text{cut-off}}} $
and
 ${\bf  M}_{J_0}^{lr}  $ is a diagonal matrix whose   ${i_1 \dots i_D}$   diagonal element    is    
   $ \left( \frac{1}{E_m-E^0_{i_1\dots i_D}} - \lambda \right) $, if ${i_1 \dots i_D} \in J_0 $   and 
zero otherwise.   
Applying    ${\bf   M}_{J_0}^{lr}        $ to a vector increases its rank by a factor of  $N_{lr}$.        Applying  
 ${\bf{  I   }}       $ to a vector does not change its rank.   Therefore   applying ${\bf M}_{J_0}$  to a vector increases its rank by a factor of   $N_{lr} + 1$. 
An ALS reduction  must be done immediately after applying ${\bf M}_{J_0}$. 
If step~(\ref{alg:precond}) is ignored, the iteration still converges because the Davidson 
space becomes identical to a Lanczos subspace (but with vectors different from those we would obtain using the Lanczos method, which does not include a  diagonalization step). 

One important advantage of the Davidson algorithm is that the selection criteria (step~\ref{select}) can take several forms. Here we will always select the eigenvectors corresponding to the lowest eigenvalues. We could also choose some overlap criteria or specify a given spectral window. 
 
The memory cost depends on the number of iterations between restarts, $N_{rs}$. 
The maximum memory cost is reached just before restarting and scales as $ \mathcal{O}(N_{rs}BTRDn) $. 
The calculation cost scales as the number of matrix-vector products, equal to 
$(2 B N_{cycle} N_{rs})$ because there are $B$ matrix-vector products for calculating the residuals and $B$ more for making or augmenting the ${\bf H}'$ matrix, for each value of the $j$ and $k$ indices. 
This has to be multiplied by the cost of one product and one reduction, Eqs. \eqref{eq:costmvp} and \eqref{eq:costals}. 
This estimate does not include the cost of applying the preconditioner which  significantly affects  the overall cost if the cut-off is large.

\subsection{Parallelization}

There are two  parallelization strategies.  Both the Davidson and the Chebyshev eigensolvers can be 
 parallelized  over vectors in the block (different vectors in a block are computed on different threads) or the  operations required to compute a single vector can 
be parallelized.   The first strategy  is easier to implement but requires more  memory because one must store many high rank vectors generated by doing MVP.
The second strategy  allows one  to store, one at a time, the  high-rank  vectors arising from  matrix-vector products. 
In practice, we usually parallelize over vectors in the block, as is done in  Ref. \cite{leclerc2014}.  
Some of the calculations have been done sequentially in order to  facilitate  CPU time comparisons. 


\section{Numerical results on the Acetonitrile molecule \label{results}}

In this section we  show that the three eigensolvers presented in section~\ref{theory} all work well and all give accurate  eigenvalues for a 12-D problem. 
In this paper,  we use the eigensolvers only in two layer calculations. Since we do not exploit the hierarchical (contraction) ideas of Ref. \cite{thomas2015}, we do
 not demand that the energies  be as accurate  as in \cite{thomas2015}. The comparison in this paper  is a relative comparison of the  different reduced-rank eigensolvers. 
If the Chebyshev and/or Davidson eigensolver is more efficient than the shifted power method  it  can be used in conjunction with the   hierarchical idea. 
The  most demanding part  of the   hierarchical  calculation is computing eigenfunctions of 
the node at the top of the tree, which typically requires many RRBPM iterations.   A more efficient eigensolver could therefore be used to improve  the H-RRBPM. 

\subsection{Hamiltonian and basis set}

We compute the acetonitrile (CH$_3$CN) vibrational spectrum. This is a six-atom molecule and we 
 calculate eigenvalues and eigenstates of a 12D quartic Hamiltonian. The normal-coordinate Hamiltonian is 
\bea
\hat{H}(q_1,\dots ,q_{12})
&=& - \frac{1}{2} \sum_{i=1}^{12}  \omega_i \frac{\partial ^2}{\partial q_i^2}
+ \frac{1}{2} \sum_{i=1}^{12} \omega_i q_i^2
+\frac{1}{6} \sum_{i=1}^{12}\sum_{j=1}^{12} \sum_{k=1}^{12} \phi^{(3)}_{ijk} q_i q_j q_k \nonumber \\
&&+\frac{1}{24} \sum_{i=1}^{12} \sum_{j=1}^{12} \sum_{k=1}^{12} \sum_{\ell=1}^{12} \phi^{(4)}_{ijk\ell} q_i q_j q_k q_{\ell},
\label{eq:Hamiltonian}
\eea
with the same assumptions as in \cite{avila2011}. In Eq. \eqref{eq:Hamiltonian}, coordinates $q_1$ to $q_4$ are non-degenerate, coordinates $q_5$ to $q_{12}$ are members of doubly-degenerate pairs. 
The potential coefficients are those used in  \cite{avila2011} and are based on the constants reported in     \cite{begue2005}. 
The direct product  basis set is a product  of 1D anharmonic eigenfunctions.
The 1D functions are obtained by diagonalizing 1D uncoupled Hamiltonians  that are obtained by setting all but one normal coordinate equal to zero.
 This preliminary calculation has been done in a  basis of harmonic oscillator eigenfunctions. The Hamiltonian operator is then factorized following \cite{thomas2015} to minimize the number of terms that need to be applied to each vector. After factorization, there are 216 terms in $\hat{H}$.

\subsection{Numerical results}

To compare the  three reduced rank eigensolvers, we  list differences between energy levels and the zero-point energy (ZPE), 
and corresponding errors 
after 20 and 100 matrix-vector products (MVP), see Table~\ref{res:differences}.  
The number of MVP is roughly proportional to the cost of the calculation.   
It is not actually the MVP itself that is costly, but the rank reduction that is done after each MVP.   
All the calculations have been done with the same  bases  (identical to  those  of \cite{avila2011,leclerc2014}), the same initial block made of $B=32$ eigenvectors
 of the separable approximation to the  Hamiltonian. We have used the same reduction rank $R=50$ for all the calculations and a fixed number of ALS iterations, $N_{ALS}=15$ for rank reductions. 
 RRBPM calculations were done with     an energy shift of $\sigma = 170000$~cm$^{-1}$ \cite{leclerc2014}. Diagonalization  and vector updates were  done every 10 iterations.
When using the  Reduced Rank Block Chebyshev (RRBC) method, diagonalizations and vector updates were done  every 10 Chebyshev iterations.  
The value of $E_{max}$ used to compute the filter parameters was padded by adding 3173.7 $\textrm{c\ensuremath{m^{-1}}}$.
The Reduced Rank Block Davidson (RRBD) method is restarted every $N_{rs}=4$ iterations to reduce  CPU  cost. 
The preconditioning step (Eq. \eqref{precond1} in section \ref{RRBDM}) is applied within an active subspace made of the first $500$ basis functions of the direct product basis set.

In the third column of table~\ref{res:differences}, we show differences between levels computed using 
the RRBD method and the corresponding ZPE,  for the first 32 vibrational states of acetonitrile.   
These results are obtained after 200 matrix-vector products.  Increasing  the number of MVP causes them to oscillate.    
The oscillations could be reduced by increasing the target rank, $R$. 
In columns 4-9 we report 
errors with respect to the Smolyak results of Ref. \cite{avila2011} for  the RRBP,  RRBC and RRBD methods.   
In the RRBPM and the RRBC  columns of table~\ref{res:differences}, differences are given  for a  fixed number of matrix-vector products (either 20 MVP or 100 MVP). 
The number of MVP is given per computed vector, i.e. we count all the MVP and divide by $B$. 
\begin{table}
\centering
\caption{First 32 levels (from which the ZPE has been subtracted)  (cm$^{-1}$) and differences with Smolyak results \cite{avila2011}, after a fixed number of matrix-vector products (mean number per computed vector). Comparison of three reduced-rank eigensolvers: reduced-rank block power method (RRBPM), reduced-rank block Chebyshev method (RRBC) and reduced-rank block Davidson method (RRBD).
The results in  column 3 are those of the Davidson calculation after 200 matrix-vector products. }  
\begin{tabular}{ccccccccc}
\hline 
Assign. & Sym. & Energy  & \multicolumn{3}{c}{Error after 20 mvp} & \multicolumn{3}{c}{Error after 100 mvp}  \\    
 &  & level  & RRBPM & RRBC & RRBD & RRBPM & RRBC & RRBD \\ 
\hline 
ZPE                  &       & 9837.498&       &  &       &      &  &      \\
$\nu_{11}$           & $E$   & 361.08  & 0.39  & 1.59 & 0.16  & 0.13 & 0.05 & 0.08 \\ 
                     &       & 361.15  & 0.48  & 1.57 & 0.18  & 0.17 & 0.02 & 0.10 \\ 
$2\nu_{11}$          & $E$   & 723.25  & 0.88  & 3.14 & 0.20  & 0.25 & 0.13 & 0.11 \\ 
                     &       & 723.63  & 1.12  & 2.75 & 0.53  & 0.57 & 0.25 & 0.36 \\ 
$2\nu_{11}$          & $A_1$ & 724.35  & 1.30  & 3.02 & 0.61  & 0.64 & 0.17 & 0.50 \\ 
$\nu_{4}$            & $A_1$ & 900.78  & 5.59  & 3.73 & 0.27  & 1.75 & 0.19 & 0.11 \\ 
$\nu_{9}$            & $E$   & 1034.40 & 9.39  & 3.92 & 0.31  & 1.79 & 0.21 & 0.19 \\ 
                     &       & 1034.74 & 10.04 & 3.84 & 0.37  & 1.84 & 0.20 & 0.23 \\ 
$3\nu_{11}$          & $A_1$ & 1087.27 & 6.28  & 2.87 & 1.10  & 2.95 & 0.49 & 0.60 \\ 
$3\nu_{11}$          & $A_2$ & 1087.40 & 6.32  & 2.69 & 1.29  & 3.10 & 0.57 & 0.64 \\ 
$3\nu_{11}$          & $E$   & 1088.55 & 6.51  & 3.25 & 1.27  & 3.30 & 0.50 & 0.52 \\ 
                     &       & 1088.63 & 6.59  & 3.06 & 1.29  & 3.50 & 0.52 & 0.73 \\ 
$\nu_{4}+\nu_{11}$   & $E$   & 1260.12 & 8.04  & 3.81 & 1.02  & 2.86 & 0.25 & 0.43 \\ 
                     &       & 1260.26 & 8.05  & 3.51 & 1.11  & 2.92 & 0.29 & 0.46 \\ 
$\nu_{3}$            & $A_1$ & 1390.79 & 5.77  & 3.85 & 1.29  & 2.45 & 0.57 & 1.05 \\ 
$\nu_{9}+\nu_{11}$   & $E$   & 1395.50 & 11.23 & 3.74 & 1.08  & 2.76 & 0.23 & 0.60 \\ 
                     &       & 1395.64 & 11.77 & 3.37 & 1.30  & 3.39 & 0.76 & 1.29 \\ 
$\nu_{9}+\nu_{11}$   & $A_2$ & 1396.46 & 11.78 & 3.40 & 1.18  & 3.21 & 0.63 & 1.76 \\ 
$\nu_{9}+\nu_{11}$   & $A_1$ & 1398.56 & 10.66 & 3.73 & 1.77  & 3.10 & 0.64 & 1.02 \\ 
$4\nu_{11}$          & $E$   & 1452.04 & 8.27  & 3.24 & 1.22  & 3.94 & 0.69 & 0.62 \\ 
                     &       & 1452.04 & 8.37  & 2.96 & 1.30  & 4.02 & 1.12 & 0.97 \\ 
$4\nu_{11}$          & $E$   & 1454.33 & 8.85  & 2.17 & 1.52  & 4.89 & 0.61 & 0.83 \\ 
                     &       & 1454.51 & 9.03  & 1.81 & 1.59  & 5.68 & 0.74 & 1.13 \\ 
$4\nu_{11}$          & $A_1$ & 1454.79 & 9.47  & 2.09 & 2.04  & 5.36 & 0.98 & 1.57 \\
$\nu_{7}$            & $E$   & 1483.57 & 5.48  & 5.39 & 0.54  & 1.14 & 0.02 & 0.20 \\ 
                     &       & 1483.61 & 6.89  & 5.38 & 0.58  & 1.59 & 0.01 & 0.26 \\ 
$\nu_{4}+2\nu_{11}$  & $E$   & 1620.57 & 11.28 & 2.75 & 1.21  & 4.26 & 0.29 & 0.45 \\ 
                     &       & 1621.41 & 11.66 & 2.35 & 1.95  & 4.87 & 0.71 & 1.23 \\ 
$\nu_{4}+2\nu_{11}$  & $A_1$ & 1622.25 & 13.26 & 1.04 & 2.49  & 6.25 & 1.32 & 1.51 \\
$\nu_{3}+\nu_{11}$   & $E$   & 1753.08 & 8.77  & 1.23 & 4.34  & 5.94 & 4.27 & 3.91 \\ 
                     &       & 1753.18 & 8.81  & 0.79 & 4.50  & 6.01 & 4.42 & 4.11 \\ 
\hline 
\end{tabular} 
\label{res:differences}
\end{table}

From table \ref{res:differences}, one concludes that all methods converge to the true eigenvalues. It is also clear that the RRBC and RRBD methods  converge 
substantially faster than the RRBPM. 
All errors of  RRBD  calculations  are smaller than their  RRBPM counterparts, 
both after 20 matrix-vector products and after 100 matrix-vector products. 
The same is true after 100 matrix-vector product for the reduced-rank Chebyshev method, 
but some errors are larger and some are smaller than those of the RRBPM if we compare  the energies  after only 20 matrix-vector products. 
It should be noted  that  errors are calculated by taking differences of energy differences and this means that the error in  the zero-point energy also contributes
 to the error estimates given in table \ref{res:differences}. 
For all three eigensolvers, the highest two energy levels  in the block  have larger errors (about 4 cm$^{-1}$).   
Except these two states, all  errors in energy differences are  between  $[0.01 , 1.32]$ cm$^{-1}$ for the Chebyshev results, $[0.08, 1.76]$ cm$^{-1}$ for the block-Davidson method whereas the RRBPM errors are between  $[0.13, 6.25]$, after 100 MVP.

\begin{figure}
\includegraphics[width=\linewidth]{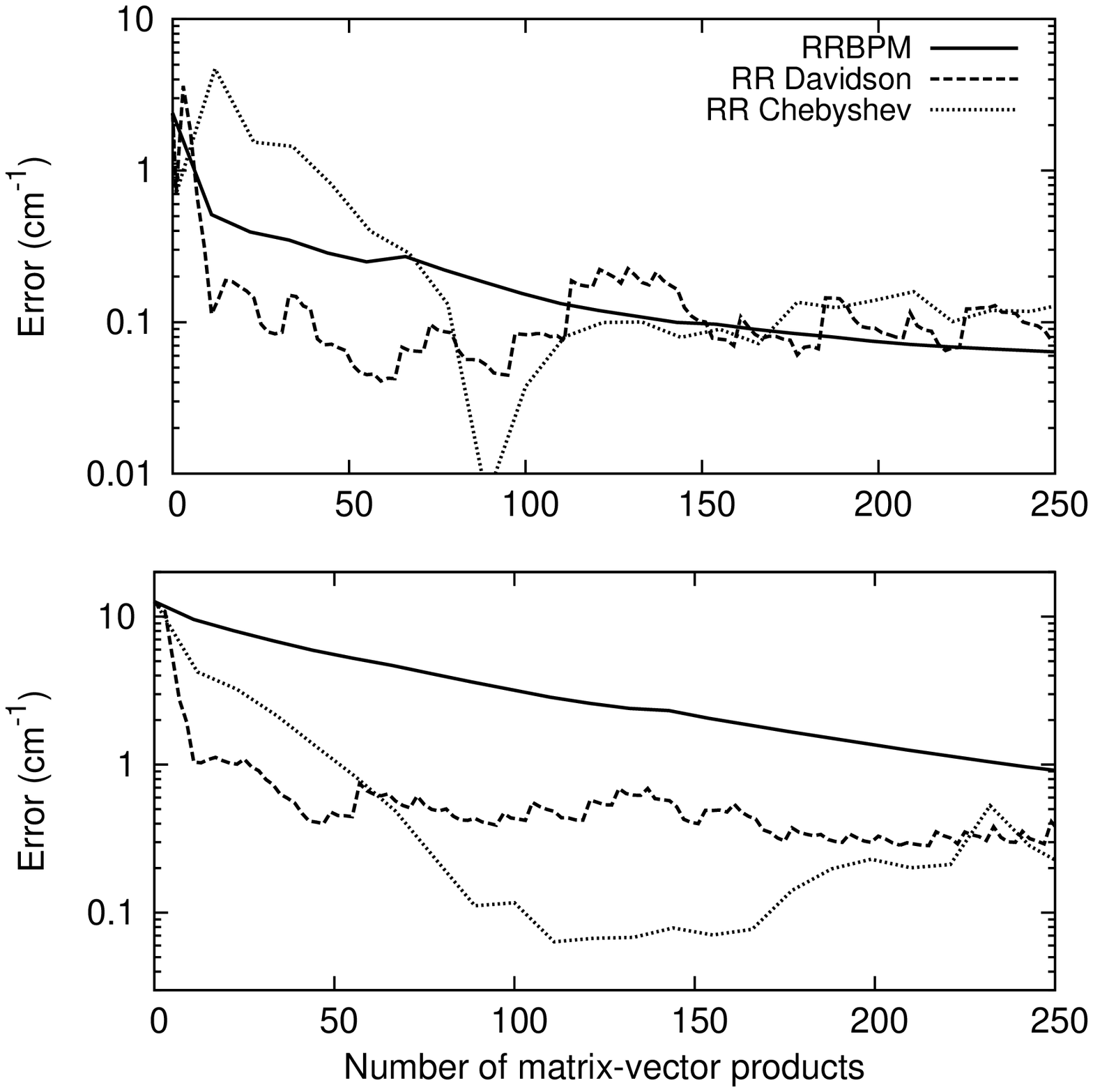} 
\caption{Convergence curves for two levels: $\nu_{11}$ (upper curves) and $\nu_{4}+\nu_{11}$ (lower curves), using RRBPM (solid line), RRBD method (dashed line) and RRBC method (dotted line). The horizontal axis is the number of matrix-vector products per computed vector and the errors are calculated with respect to results of \cite{avila2011}.} 
\label{fig1}
\end{figure}

Convergence curves for two representative 
levels, the first excited state $\nu_{11}$ and  the $13^{th}$ state $\nu_{4}+\nu_{11}$,  are given in Fig.   \ref{fig1}.   
Convergence curves for  the power method are almost monotonic but decreases slowly. 
The Davidson results exhibit first a global descent but then small oscillations. The Davidson oscillations occur because error increases 
when the algorithm is restarted with         a new, smaller block of vectors,
and decreases as number of vectors in space again increases.   
The first few MVP decrease the  RRBD error  very quickly.

The RRBC algorithm converges much more quickly than the RRBPM during the first tens of iterations for all energy levels considered.
However, later on oscillations set in, due to the imposed reduction rank. For the RRBPM, the oscillations do not appear for all levels,
but for  the levels that do oscillate  they are typically less  than 0.01 cm$^{-1}$ in magnitude. 
For the RRBC algorithm the oscillations are larger, having magnitudes of $\sim$0.1 cm$^{-1}$ for all levels. 
For both  solvers  the oscillations can be dampened by increasing the rank.

\begin{table}
\centering
\caption{Speed-up ratios with respect to RRBPM \cite{leclerc2014}, based on the CPU time required to achieve frequencies within 2 cm$^{-1}$ of the Smolyak values, for the first 10, 20, 25 levels  above the ground state, for the reduced rank Chebyshev and Davidson methods described in section \ref{theory}. The last column is obtained by omitting the orthogonalization step in the Davidson algorithm. The speed-up ratios are computed using CPU times from sequential calculations. }
\begin{tabular}{cccc}
\hline 
Number of converged & $\frac{T_{cpu}(\text{RRCheb})}{T_{cpu}(\text{RRBPM})}$ & $\frac{T_{cpu}(\text{RRBDav.})}{T_{cpu}(\text{RRBPM})}$ & $\frac{T_{cpu}(\text{RRBDav.})}{T_{cpu}(\text{RRBPM})}$ \\ 
levels  & & & no orthog. \\
\hline 
first 10   & 0.295 & 0.249 & 0.217 \\ 
first 20   & 0.218 & 0.293 & 0.287 \\ 
first 25   & 0.180 & 0.321 & 0.279 \\ 
\hline 
\end{tabular} 
\label{res:ratios}
\end{table}

In the RRBD method  there are also orthogonalization and preconditioning steps which do not influence the number of MVP but do increase  the CPU cost.
 Therefore, it is also important to compare the cost of the three calculations. 
In table \ref{res:ratios}, 
we report  ratios of  RRBC and RRBD  CPU times to the RRBPM CPU time.  
 These  speed-up ratios are  computed using the CPU time, with no parallelization,  required to achieve a difference of less than 2 cm$^{-1}$ with respect to Smolyak results  for the first ten, for the first twenty, and for the first twenty-five  energy levels.  
The RRBD  method reduces the CPU cost by factor of 3-4 for all three groups of states. 
The CPU time ratios could be further   improved by omitting the orthogonalization step and the associated reduction (step \ref{step:orth} of the Davidson algorithm in subsection \ref{RRBDM}), which are not essential because we diagonalize using a generalized eigenvalue algorithm anyway. 
The RRBC          method is a bit more efficient than the Davidson method, the former being 4-5 times faster than the RRBPM. The largest 
 speed-up ratio is obtained when comparing convergence for larger numbers of eigenvalues because the RRBPM generally fails to give the largest eigenvalues in the wanted block. 
We conclude that both reduced rank iterative methods presented in section \ref{theory} converge faster than the RRBPM with similar speed-up ratios, with the Chebyshev method being slightly faster. 

All three eigensolvers  require little memory and the memory cost scales linearly with dimensionality. 
The  RRBD  method  has the highest memory cost,   since the subspace size increases during the calculation.
 However, due to the use   of  CP format, the  memory requirement is so low that it is not  important. 
One reduced CP-vector with rank $50$ takes $55$ kB. After multiplication by ${\bf H}$, the rank becomes temporarily larger (approximately 10000) with a memory cost of $12$ MB per vector.


\section{Conclusion \label{conclusion}}


The memory cost of variational calculations has limited them to molecules with fewer than about six atoms.   Modern methods   all use  an iterative algorithm, based on 
evaluating matrix-vector products,  to compute 
eigenvalues and eigenvectors and require storing only a few vectors  in memory.   Nonetheless, if a simple direct product basis is used the memory cost of such calculations 
is prohibitive because each vector has $n^D$ components. 
Although it is possible to avoid direct product bases, they have the advantage of being simple and easy to use.   It is therefore important to 
explore ideas that make it possible to use a direct product basis without storing   $n^D$ numbers.    For potentials in SOP form this is possible if one uses  an 
iterative eigensolver in conjunction with  tensor rank reduction.   One uses  an iterative eigensolver to generate basis vectors, reduces their rank, and then computes eigenvalues
by projecting into the space spanned by the reduced basis vectors  \cite{leclerc2014}. Many variants of this idea are possible \cite{rakhuba2016}.   It is clear that the 
block power method used in \Ref{leclerc2014} is not optimal.  
 In this paper we have assessed the 
advantages of two other iterative eigensolvers. Both the Cheybshev and the Davidson methods significantly reduce the CPU time.  
The memory cost  of the RRBD method  is  greater than that of the RRBPM of Ref. \cite{leclerc2014} due to the growth of  subspace.  
The memory cost of the RRBC method is essentially the same as that of  the RRBPM.
However, the memory cost is very low (and scales linearly with $D$)  compared to that of standard iterative direct product calculation.
The RRBPM remains the simplest method to implement. 
Improved eigensolvers can be coupled with the 
hierarchical ideas of \cite{thomas2015} that use successive contractions and intermediate diagonalizations.

\section*{Acknowledgments}
Some  of the calculations were done   on computers purchased with a grant for the Canada Foundation for Innovation.  This research was funded by the 
Natural Sciences and Engineering Research Council of Canada.  The PMMS (P\^ole Messin de Mod\'elisation et de Simulation) is gratefully acknowledged for providing us with computer time.

\bibliographystyle{elsarticle-num}
\bibliography{rreigensolvers}

\end{document}